\documentclass[pra,twocolumn,groupedaddress,shownopacs,amssymb]{revtex4-1}
\usepackage{graphicx,psfrag,amsmath,amssymb}
\usepackage[usenames, dvipsnames]{color}
\usepackage{braket}
\usepackage{amsmath}
\usepackage{amssymb}
\usepackage{float}
\usepackage{gensymb}
\usepackage{makeidx}
\usepackage{mathtools}
 
\usepackage[normalem]{ulem}
\usepackage[T1]{fontenc} 
\usepackage{balance}
\usepackage{flushend}
\usepackage{lineno}
\usepackage{epsfig}
\usepackage{subfigure}
\usepackage{mathtools}
\usepackage{setspace}
\usepackage{bm}
\usepackage{ulem}
\usepackage{times}
\usepackage{enumitem}
\usepackage{xcolor}
\usepackage{multirow}
\usepackage{soul}
\usepackage{dcolumn}
\usepackage{balance}
\usepackage{epstopdf}
\usepackage{braket}
\usepackage{color}
\usepackage{diagbox}
\usepackage[utf8]{inputenc}
\usepackage{hyperref}
\hypersetup{
    colorlinks=true,
    linkcolor=Blue,
    citecolor=Blue,
    filecolor=Blue,
    urlcolor=Blue,
    }
\begin{document}
\title{Potential-defect-driven collective modes of one-dimensional two-component quantum droplets}
\author{Harsimranjit Kaur}
\affiliation{Department of Physics, Central University of Rajasthan, Ajmer - 305817, India}
\author{Kuldeep Suthar}
\affiliation{Department of Physics, Central University of Rajasthan, Ajmer - 305817, India}

\date{\today}
\begin{abstract}
 We examine a one-dimensional binary mixture of ultradilute quantum droplets in the presence of a central potential defect. The properties of the potential strength and the number of atoms exhibit distinct polarization transitions and low-lying collective excitation spectra. The balance of (attractive) Lee-Huang-Yang quantum fluctuations and repulsive mean-field interactions results in a self-bound quantum droplet in a potential well. However, the potential barrier causes the polarization transition with the number of atoms, which is reflected in the excitation spectrum as a discontinuity and softening of the quasiparticle modes. We reveal that the critical number of atoms for the transition decreases as the attractive intercomponent interaction increases. Finally, the quench time dynamics of the interaction in potential barrier and well show the localization and diffusive fragmented droplets of two-component systems.  
\end{abstract}
\maketitle
\section{Introduction}\label{Introduction}
 The many-body ultradilute quantum droplet state represents a novel self-bound state of matter stabilized by the interplay between attractive mean-field interactions and beyond-mean-field Lee-Huang-Yang (LHY) quantum fluctuations, as predicted first in Ref.~\cite{petrov_15} and subsequently observed in homonuclear and heteronuclear Bose-Bose mixtures
~\cite{cabrera_18,cheiney_18,ferioli_19,DErrico_19,Burchianti_20}, as well as in long-range interacting dipolar gases~\cite{bottcher_21,Chomaz_23,Barbut_16,Schmitt_16}. 
The LHY correction stabilizes the formation of a nontrivial quantum droplet state~\cite{Luo_21,malomed_20,wachtler_16} and depends on interatomic interactions and the dimensionality of the system~\cite{lee_57,petrov_16,ilg_18}. Although the correction is smaller than the mean-field contributions, the tunability of the interaction in quantum gas experiments~\cite{Chomaz_23,Chen_21} allows an ideal platform to control the mean-field effects. This is facilitated by recent realizations of the soliton--droplet crossover~\cite{cheiney_18,cappellaro_18,semeghini_18,cabrera_18} and liquid-like properties, including collision dynamics governed by surface tension and self-binding~\cite{ferioli_19,petrov_16,wachtler_16}. The low-lying collective excitations such as breathing and surface modes associated with finite compressibility and broken symmetries~\cite{tylutki_20,sturmer_21,fei_24,zin_21,Baillie_16}, provide key insights into droplet stability and dynamical response. The external confining potential plays a crucial role in trapping the atoms; experimental advances in controlling the nature of the trap provide avenue to explore the self-binding of quantum droplet states under various potential perturbations, resulting in collective oscillations and tunneling phenomena~\cite{albiez_05,englezos_25,Pelayo_25,smerzi_97,levy_07,wachtler_16}. 

Localized central potential defects, such as dimple and repulsive barrier potentials, provide an ideal platform for controlling and modifying the collective properties of atomic
condensates~\cite{shin_04,schumm_05,jo_07,uncu_07,garrett_11}. In particular, trapping geometries such as optical lattice microtraps~\cite{morsch_06,bloch_05}, box
potential~\cite{gaunt_13,navon_16}, localized Gaussian barrier~\cite{kaur_25,levy_07,shin_04}, and dimple potential~\cite{garrett_11,uncu_07,ramanathan_11} have been extensively used to
understand stationary and dynamical properties. Double-well potentials realized via adiabatic radio-frequency–induced splitting enable precise control over coherent splitting and recombination in quantum dynamics~\cite{schumm_05,jo_07,shin_04}, facilitating detailed investigation of phase coherence and tunneling phenomena. The population imbalance and relative phase in the wells determine the fate of Josephson oscillations and macroscopic quantum self-trapping~\cite{albiez_05,levy_07,jo_07}, where earlier is analogous to the Josephson effect in
superconductors~\cite{smerzi_97,raghavan_99,albiez_05,levy_07}. More recently, multiple quantum droplets of heteronuclear quantum mixtures and fragmentation phenomena have been realized by quenching the interaction into a strongly attractive regime~\cite{cavicchioli_25}. Despite significant progress in the study of quantum droplets, the effects of defect potentials on the ground state (GS) properties and collective excitations remain relatively underexplored.

In this work, we study the ground-state properties and low-lying collective excitation spectra of a two-component quantum droplet in the presence of a localized central potential barrier or well, by varying the number of atoms and controlling the harmonic confinement. We employ the extended Gross-Pitaevskii equation (eGPE) with a beyond-mean-field LHY correction and Bogoliubov theory~\cite{pitaevskii_16,englezos_25} to obtain the stationary properties and quasiparticle energy of collective modes. The potential well causes a transition from Gaussian density to flat-top distribution that becomes narrower around the center of the trap. The presence of a barrier induces a transition from a polarized to an unpolarized state with the number of atoms, which is reflected in the excitation spectrum as a discontinuity in the quasiparticle spectra. The critical number of atoms for the transition to occur decreases as the intercomponent interaction becomes less attractive. The dynamical response of the droplet to quench in the intercomponent attractive interaction and defect exhibit breathing oscillations, solitary wave generation, and fragmentation of droplet.

This paper is structured as follows. In Sec.~\ref{THEORY}, we introduce the extended coupled Gross–Pitaevskii equations and Bogoliubov theory as the basis of our investigation. In Sec.~\ref{RESULTS AND DISCUSSIONS}, we discuss the polarization transition with the number of atoms in a repulsive barrier and low-lying collective quasiparticle modes of two-component quantum droplet. We further examine the temporal dynamics of droplet for different quench of intercomponent interaction and potential. Finally, we conclude in Sec.~\ref{conclusion}.

\section{THEORY}\label{THEORY}
Consider a homonuclear binary bosonic mixture which is tightly confined in the transverse directions, and thus the atomic motion is effectively frozen along the perpendicular plane. The quantum droplet mixture in such a quasi-one-dimensional (along $x$-axis) system is achieved for repulsive intracomponent and attractive intercomponent interaction strengths. The two-component quantum droplet mixture has been realized in recent experiments~\cite{semeghini_18,cabrera_18,ferioli_19} for \(^{39}\mathrm{K}\) in the hyperfine states $\ket{F=1, m_{F}=-1}$ and $\ket{F=1, m_{F}=0}$. By incorporating beyond-mean-field Lee--Huang--Yang correction, the energy density of droplet mixture is written as~\cite{petrov_15,wachtler_16} 
\begin{equation}
\begin{aligned}
\mathcal{E}[\Psi_{1},\Psi_{2}]
=
\int dx \Bigg[
&\begin{pmatrix}
\Psi_{1}^\dagger & \Psi_{2}^\dagger
\end{pmatrix}
\hat{h}_0
\begin{pmatrix}
\Psi_{1} \\
\Psi_{2}
\end{pmatrix}
+ \frac{g_{11}}{2} |\Psi_{1}|^4
 \\
& + \frac{g_{22}}{2} |\Psi_{2}|^4+ g_{12} |\Psi_{1}|^2 |\Psi_{2}|^2
+ \mathcal{E}_{\mathrm{LHY}}
\Bigg],
\end{aligned}
\label{en_den}
\end{equation}
where $\Psi_{1}$ and $\Psi_{2}$ are the complex bosonic fields of the first and second components. Here, $g_{11,22}$ and $g_{12}$ represent the strength of repulsive intracomponent and attractive intercomponent interactions. The single-particle Hamiltonian is $\hat{h}_0 = -(\hbar^2/2m) \partial^2/\partial x^2 + V(x)$ with $m$ being the atomic mass of each component. The external potential reads
\begin{equation}
V(x)
=
\frac{1}{2} m \omega_{x}^2 x^2
+
V_0 \exp\!\left(-\frac{x^2}{2\sigma^2}\right),
\end{equation}
which consists of a harmonic trap with trapping frequency $\omega_{x}$ with $\omega_{x}\ll\omega_{\perp}$ for the quasi-1D geometry. The trapping frequency is $\omega_{x} = \lambda\omega_{o}$ \cite{xiao_26} where $\lambda$ is the strength of the one-dimensional harmonic trapping potential. The second term of $V(x)$ is a localized Gaussian defect of strength $V_0$ and width $\sigma$. The nature of the strength determines the effective potential to be a barrier or a potential well. The last term in the energy density [Eq.~\eqref{en_den}] is a beyond-mean-field LHY correction which is given as~\cite{englezos_25,mistakidis_23},
\[
{\mathcal{E}}_{\mathrm{LHY}}
=
\frac{\sqrt{m}}{2\pi\hbar}
\left(
g_{11} n_{1}
+
g_{22} n_{2}
\right)^{\frac{3}{2}}
\mathcal{F}(p),
\]
with the function is 
\[
\mathcal{F}(p)
=
-\frac{\sqrt{2}}{3}
\left[
\left( 1 - \sqrt{p+1} \right)^{\frac{3}{2}}
+
\left( 1 + \sqrt{p+1} \right)^{\frac{3}{2}}
\right],
\]
where $p = 4(g^{2}_{12} - g_{11}g_{22})n_{1}n_{2}/(g_{11}n_{1} + g_{22}n_{2})^{2}$. The energy density mentioned above [Eq.~\eqref{en_den}] is valid for any value of the interaction strength and the number of atoms in the mean-field limit. The LHY energy density is real and negative for $g_{12} \leqslant \sqrt{g_{11}g_{22}}$. This interaction range includes both the droplet threshold ($g_{12}\approx-\sqrt{g_{11}g_{22}}$) and the immiscibility condition $g_{12}\approx\sqrt{g_{11}g_{22}}$. It is important to note that in the limit of $p=0$, the function takes the value $\mathcal{F}(0) = -4/3$, which leads to a usual contribution of the one-dimensional energy density of the droplet suggested by Petrov~\cite{petrov_16}. For an uncoupled mixture with $g_{12}=0$, $p$ becomes $-1$ and the function $\mathcal{F}(-1) = -2\sqrt{2}/3$, hence the contribution of LHY is reduced by a factor of $\sqrt{2}$. The exact correction allows us to examine the collective modes away from the mean-field stability regime. It has recently been considered in recent studies of quantum droplets~\cite{bristy_25,englezos_25} and mobile impurities~\cite{sinha_23}.

The Euler-Lagrange or Gross-Pitaevskii (GP) equations describing the dynamics of the system are
\begin{equation}
\begin{aligned}
i\hbar\frac{\partial \psi_i}{\partial t}&=\Bigg[-\frac{\hbar^2}{2m}\frac{\partial^2}{\partial x^2}+V(x)+g_{ii}|\psi_i|^2+g_{i3-i}|\psi_{3-i}|^2\Bigg]\psi_i\\ &+\frac{\partial\mathcal{E}_{\rm LHY}}{\partial n_i}\psi_i. 
\label{egpe_equation_2}
\end{aligned}
\end{equation}
Here,
\begin{equation}
\begin{split}
\frac{\partial \mathcal{E}_{\mathrm{LHY}}}{\partial n_i}
&=
\frac{3g_{ii}}{4\pi\hbar}
\sqrt{m(g_{ii}|\psi_i|^2+g_{i3-i}|\psi_{3-i}|^2)}
\,\mathcal{F}(p) \\
&\quad+
\frac{\sqrt{m\left(g_{ii}n_1+g_{i3-i}n_2\right)^3}}
{2\pi\hbar}
\frac{\partial\mathcal{F}}{\partial p}
\frac{\partial p}{\partial n_i},
\end{split}
\label{lhy_term}
\end{equation}
where \[\frac{\partial\mathcal{F}}{\partial p}=\frac{\sqrt{2}(\sqrt{1-\sqrt{p+1}}-\sqrt{1+\sqrt{p+1}}}{4\sqrt{p+1}}\] and \[\frac{\partial p}{\partial n_i}=\frac{4n_{3-i}(g_{ii}g_{3-i3-i}-g_{i3-i}^2)(g_{ii}n_i-g_{3-i3-i}n_{3-i})}{(g_{ii}n_1 +g_{3-i3-i}n_{2})^3}\]
Here, $i=1,2$ is the component index. The second term of Eq.~\eqref{lhy_term} vanishes by considering the limiting case of a fully balanced mixture i.e. $g_{11}=g_{22}=g$ and $n_1 =n_2$, which makes $p=(g_{12}^2/g-1)$.
The coupled extended Gross-Pitaevskii equation in dimensionless form reads as
\begin{equation}
\begin{aligned}
i\frac{\partial \psi_i}{\partial t}&=\Bigg[-\frac{1}{2}\frac{\partial^2}{\partial x^2}+V(x)+g|\psi_i|^2+g_{i3-i}|\psi_{3-i}|^2\Bigg]\psi_i\\ &+\frac{3g^{3/2}}{4\pi}\sqrt{|\psi_i|^2+|\psi_{3-i}|^2}\,\mathcal{F}(p)\,\psi_i. 
\label{egpe_equation_2}
\end{aligned}
\end{equation}
The energy and potential strength are expressed in units of characteristic energy $\hbar\omega_{o}$, length in $\sqrt{\hbar/(m\omega_{o})}$, and time in $\omega_{o}^{-1}$. The wavefunction of the $i$th component $(i=1,2)$ satisfies the normalization condition $N_i = \int |\psi_i(x)|^2 dx$ that acts as an effective controlling parameter of the number of atoms.

To investigate the collective excitations in a weakly-interacting Bose gas, the Bogoliubov theory or linearization technique is used. The time-dependent Bose field can be written in terms of the static ground state and small-amplitude fluctuations around the ground state stemming from the presence of quantum fluctuations. The Bose field of eGPEs can be linearized as~\cite{bogoliubov_47,pitaevskii_16} 
\begin{align}
\psi_i(x,t)=e^{-i\mu_i t}\left[\phi_i(x)+\sum_j\{u_{ij}(x)e^{-i\omega_j t}+v_{ij}^{*}(x)e^{i\omega_j t}
\}
\right],
\end{align}
where $\phi_i(x)$ is the stationary ground-state wave function of the $i$th component, while $u_{ij}(x)$ and $v_{ij}(x)$ denote the Bogoliubov quasiparticle amplitudes corresponding to the $j$th mode excitation with quasiparticle frequency $\omega_j$. Here, $\mu_i$ is the chemical potential of the $i$th component. Substituting the above approximation in Eq.~\eqref{egpe_equation_2} and retaining only linear terms in the quasiparticle amplitudes yields the Bogoliubov-de Gennes (BdG) eigenvalue equations~\cite{fetter_72,mistakidis_23}
\begin{equation}
\omega_j
\begin{pmatrix}
u_{1}\\
v_{1}\\
u_{2}\\
v_{2}
\end{pmatrix}
= \mathcal{H}_{\mathrm{BdG}}
\begin{pmatrix}
u_{1}\\
v_{1}\\
u_{2}\\
v_{2}
\end{pmatrix}.
\end{equation}
\begin{figure*}[hbt]
  \includegraphics[width=\textwidth]{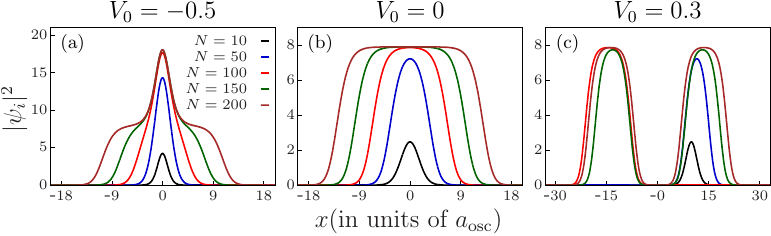}
  \caption{The ground state density profiles of bosonic quantum droplet mixtures for different numbers of atoms at three barrier strengths: (a) $V_0=-0.5$ (attractive defect), (b) $V_0=0$ (no defect), and (c) $V_0=0.3$ (repulsive defect). The intercomponent interaction strength is fixed at $g_{12}=-0.9$, and the harmonic trap strength is set to $\lambda=0$. The quantum states are stabilized due to LHY correction. Different lines present the density profiles for different $N$ labeled in (a).}
\label{density_w_natoms}
\end{figure*}
The BdG Hamiltonian $\mathcal{H}_{\mathrm{BdG}}$ can be written in matrix form as
\begin{equation}
\mathcal{H}_{\mathrm{BdG}} =
\begin{pmatrix}
\mathcal{A} & \mathcal{B} & \mathcal{C} & \mathcal{D} \\
- \mathcal{B}^{*} & -\mathcal{A}^{*} & -\mathcal{D}^{*} & -\mathcal{C}^{*} \\
\mathcal{C}^{*} & \mathcal{D} & \mathcal{H} & \mathcal{G} \\
-\mathcal{D}^{*} & -\mathcal{C}^{*} & -\mathcal{G}^{*} & -\mathcal{H}^{*}
\end{pmatrix},
\label{bdg_matrix}
\end{equation}
with the matrix elements read as
\begin{align*}
\mathcal{A} =&-\frac{1}{2}\partial_x^2 + V(x) - \mu_{1}+ 2 g|\phi_{1}|^2 +  g_{12}|\phi_{2}|^2 \\
&+ \alpha g\left( 3  |\phi_{1}|^2 + 2   |\phi_{2}|^2 \right), \\
\mathcal{H} =&-\frac{1}{2}\partial_x^2 + V(x) - \mu_{2}+ 2 g|\phi_{2}|^2 + g_{12}|\phi_{1}|^2 \\
&+ \alpha g\left( 3 |\phi_{2}|^2 + 2  |\phi_{1}|^2 \right) \\
\mathcal{B} =& \left(1 + \alpha  \right) g \phi_{1}^{2},
\qquad
\mathcal{G} = \left(1 + \alpha  \right) g \phi_{2}^{2}, \\
\mathcal{C} =& \left(1 + \alpha \right) g_{12} \phi_{1} \phi_{2}^{*}, 
\qquad
\mathcal{D} = \left(1 + \alpha \right) g_{12} \phi_{1} \phi_{2}.
\end{align*}
Here, the coefficient $\alpha$ arises from the beyond-mean-field contribution and is given by
\begin{equation*}
	\alpha =\frac{3 \sqrt{g} \mathcal{F}(p)}{8\pi \sqrt{ |\phi_{1}|^2 +  |\phi_{2}|^2}}.
\end{equation*}
The eigenvalues of the coupled BdG equations correspond to the quasiparticle excitation frequencies of the system, while the eigenstates are the quasiparticle amplitudes $(u_i, v_i)$ that characterize the nature of the excitations~\cite{fetter_72,petrov_23,englezos_25}. Here, ground-state solutions $\phi_i$ are obtained numerically using the imaginary-time propagation of eGPEs, and the self-consistent diagonalization of coupled BdG equations gives the quasiparticle excitations of bosonic quantum droplet mixtures.

\section{Results and Discussions}\label{RESULTS AND DISCUSSIONS}
We examine the stationary properties including the excitation spectrum and dynamical properties of a one-dimensional binary Bose mixture confined in a tunable harmonic trap with localized potential defect. The stationary state is obtained by solving the coupled extended Gross--Pitaevskii equations [Eq.~\eqref{egpe_equation_2}] with a random initial state and a superposition of symmetric and asymmetric functions. Collective excitations are determined by diagonalizing the Bogoliubov-de Gennes matrix in Eq.~\eqref{bdg_matrix} using the basis of $200$ harmonic-oscillator states. The bosonic mixture consists of equal mass of each component, repulsive intracomponent interactions with $g=1$, and attractive intercomponent interactions $g_{12} < 0$ which facilitate the formation of 1D droplet configurations. The system is confined in a highly anisotropic harmonic trap, where $\omega_\perp/(2\pi) = 1\,\mathrm{kHz}$ and $\omega_x/(2\pi) = 1\,\mathrm{Hz}$, consistent with recent experimental parameters \cite{gorlitz_01,moritz_03,romero_24}. The localized defect potential at the center, which can act as either a repulsive barrier or an attractive well, can be realized using a focused laser beam, thereby allowing precise control over the ground-state, collective excitations, and dynamical behavior of the system~\cite{gaunt_13,navon_21,tajik_19}.
\subsection{Stationary properties}
\begin{figure}[h]
  \includegraphics[width=0.48\textwidth]{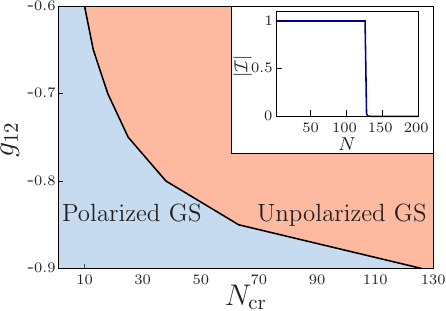}
  \caption{The critical atom number $N_{\mathrm{cr}}$ for different attractive intercomponent interaction strength $g_{12}$. At $ N < N_{\mathrm{cr}}$, the ground state becomes highly polarized, 
           indicating localization of the condensate in one side of the double well. The inset shows the population imbalance as a function of the number of atoms for $V_0 = 0.3$ and $g_{12} = -0.9$. Here, the barrier width is $\sigma=1$.}
  \label{g12_vs_ncr}
\end{figure}
We first discuss the ground state of a 1D droplet in the presence of a localized potential defect. Fig.~\ref{density_w_natoms} presents the density profiles for different numbers of atoms in the absence of harmonic confinement with $\lambda=0$. Fig.~\ref{density_w_natoms}(a) shows the densities with an attractive barrier ($V_0<0$). The external defect develops a central peak within the trap; in particular, for a large number of atoms ($N=200$), the ground state is strongly localized and narrowly confined condensate around the center of the potential. As the number of atoms decreases, the combined effect of reduced mean-field interactions and weaker LHY corrections diminishes the influence of the central peak, resulting in a less localized density profile, as observed for $N=10$. For a smaller number of atoms, a plateau appears near the central peak before the density approaches zero at the edges. When the barrier is removed, the density profiles exhibit the formation of a quantum droplet~\cite{fei_24,petrov_16,astrakharchik_18,du_23}, as shown in Fig~\ref{density_w_natoms}(b). This is evident from the transition in density from the Gaussian distribution for a smaller $N$ to a flat-top structure with a saturated central density for a larger $N$, beyond a critical $N$. It is worth noting that the stable and self-bound quantum droplet emerges from the delicate balance between repulsive mean-field interatomic interactions and attractive quantum fluctuations of the LHY correction.
\begin{figure}[h]
 \includegraphics[width=0.48\textwidth]{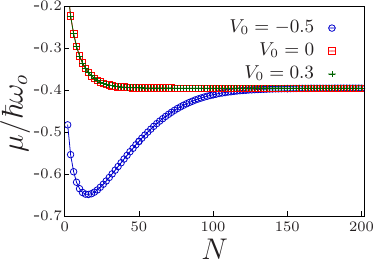}
   \caption{The chemical potential $\mu(N)$ for different barrier strengths $V_0$, with both $\mu$ and $V_0$ measured in units of $\hbar\omega_{o}$. The saturation of $\mu$ beyond the critical atom   
            number $N_{\mathrm{cr}}$ marks the flat-top droplet regime. Here, the $g_{12} = -0.9$. }
  \label{chemical potential}
\end{figure}
The presence of a repulsive defect ($V_0>0$) hosts a qualitatively different transition in the ground state. As shown in Fig.~\ref{density_w_natoms}(c), at smaller numbers of atoms, the ground state is preferentially localized in only one of the wells of the double-well potential and thus becomes strongly polarized. In this regime, the overall interaction energy is relatively weak, and the barrier-induced energy asymmetry dominates, making the symmetric (unpolarized) configuration energetically unfavorable. As a result, the condensate spontaneously breaks the symmetry and occupies only one side of the trap. The number of atoms enhances the interatomic repulsion energy  and eventually overcomes the barrier-induced localization phenomenon. We find that at critical $N_{\mathrm{cr}} = 126$ (for $g_{12} = -0.9$), the system restores the unpolarized ground state, where the condensate density becomes symmetric in both wells~\cite{trenkwalder_16}. Thus, the polarization transition of the repulsive barrier is determined by the interaction-driven delocalization and the barrier-induced localization.

In order to ascertain the polarization transition, we examine the effects of the critical number of atoms on $g_{12}$. Fig.~\ref{g12_vs_ncr} illustrates the variation of $N_{\mathrm{cr}}$, which characterizes the polarization transition, on the attractive intercomponent interaction strength $g_{12}$ in the presence of a repulsive barrier strength $V_0=0.3$. For $N < N_{\mathrm{cr}}$, the effective interaction in the system is relatively weak and the barrier-induced localization dominates, resulting in a polarized ground state. When $N > N_{\mathrm{cr}}$, the interaction energy exceeds the localization energy imposed by the barrier, thereby restoring an unpolarized (symmetric) ground state~\cite{wysocki_24}. In particular, with less attractive $g_{12}$, $N_{\mathrm{cr}}$ decreases. This is attributed to the attraction that inhibits the transition of atoms from one well to the other to minimize the ground state energy. The polarization transition can be measured by the population imbalance between the two wells, as defined by~\cite{smerzi_97,julia_diaz_09}
\begin{equation}
{\mathcal{I}}(t) = \frac{N_L(t) - N_R(t)}{N_L(t) + N_R(t)},
\end{equation}
with the number of atoms in the left and right wells given by $N_L(t) = \int_{-\infty}^{0} |\psi(x,t)|^2\, dx$ and $N_R(t) = \int_{0}^{\infty} |\psi(x,t)|^2\, dx$. The inset of Fig.~\ref{g12_vs_ncr} shows the imbalance for $g_{12}=-0.9$. The imbalance in population in the ground state is unity for $N < N_{\mathrm{cr}}$, indicating that the condensate is fully localized on one side of the double well. For $N \geqslant N_{\mathrm{cr}}$, $\mathcal{I}$ becomes zero, reflecting a symmetric distribution of the condensate across the wells.
\begin{figure}[H]
  \includegraphics[width=0.48\textwidth]{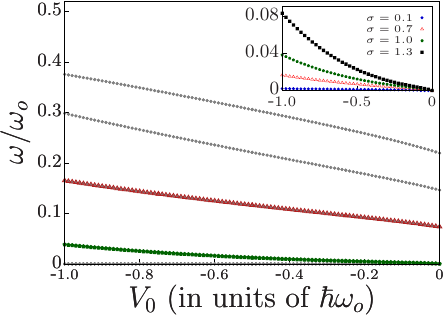} 
  \caption{The frequency of low-lying collective modes as the potential defect is tuned from attractive to zero. The lowest zero-energy modes are the Goldstone modes corresponding to each 
           component. The inset shows the enlarged view of the softening of (first excited) dipole mode with $V_0$ for different values of the spatial width $\sigma$ of potential. Here, the number of atoms is $N=150$ and main figure corresponds to $\sigma=1$. The quasiparticle frequency is scaled with trapping frequency $\omega_{o}$ and potential strength with corresponding quantum energy scale $\hbar\omega_{o}$.}
    \label{mode_with_V}
\end{figure}
\begin{figure*}[hbt]
\includegraphics[width=\textwidth]{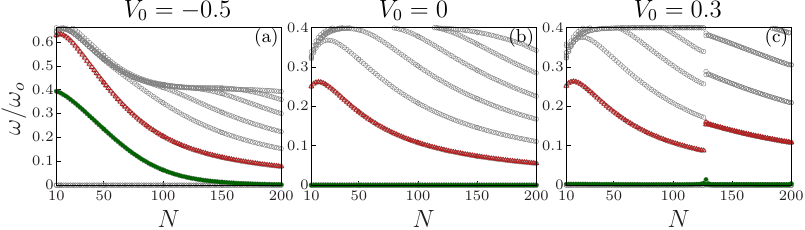}
\caption{Low-lying collective excitation frequencies as a function of the number of atoms for three different barrier strengths: (a) $V_0=-0.5$, (b) $V_0=0$, and (c) $V_0=0.3$. The intercomponent interaction strength is fixed at $g_{12}=-0.9$, and the harmonic trap strength is set to $\lambda=0$. The green filled stars and red triangles denote the frequency of dipole and breathing modes, respectively, while the light-gray dots represent the higher mode frequencies.}
\label{mode_w_natoms}
\end{figure*}
We further characterize the formation of the quantum droplet by analyzing the chemical potential at different barrier strengths. Fig.~\ref{chemical potential} presents the change in chemical potential 
$\mu$ with the number of atoms $N$ for three different values of potential strengths. For $V_0 = -0.5$, which corresponds to a central dimple potential, the chemical potential initially decreases rapidly with $N$. This is attributed to the dominating attractive contribution (or effective softening of the confinement) on the interaction energy, leading to a lowering of the total energy per particle. With further increase in $N$, $\mu$ reaches a minimum, signaling a crossover where the repulsive mean-field interaction energy begins to balance out by the attractive LHY contribution. Beyond this value of $N$, the chemical potential gradually increases and tends toward a steady value, indicating that the system approaches an interaction-dominated regime in which the density profile becomes more stabilized. For $V_0 = 0$, the chemical potential decreases initially with increasing $N$, reaches a minimum, and then saturates with a larger number of atoms~\cite{du_23}. Since the contributions of the trap to $\mu$ is very small, the case with $V_0 = 0.3$ qualitatively follows a similar trend. However, the depth of the minimum and the rate of variation depend on $V_0$. 
\subsection{Collective excitations}
We now examine the low-lying collective modes by tuning the central potential defect from the attractive regime to approach zero with attractive $g_{12}$. The evolution of low-lying quasiparticle frequencies as a function of potential strength $V_{0}$ is presented in Fig.~\ref{mode_with_V} for $N = 150$ and $\sigma=1$. The two zero-energy modes correspond to the global $U(1)$ phase invariance of two-component system, which is associated with an infinitesimal global phase shift of the coupled component wave functions. As the depth of the potential well approaches zero, the first nonzero frequency mode (dipole mode) associated with the center-of-mass motion of the droplet softens and becomes zero at $ V_{0}=0$, resulting in an additional zero-energy Goldstone mode due to the restoration of translational invariance~\cite{zin_21}. 
It is important to note that the harmonically trapped 1D condensate exhibits dipole mode oscillates with oscillator frequency, as per Kohn's theorem, while for the present case, in the absence of any external confinement, the self-bound droplet state contributes to additional zero-energy mode. Furthermore, the frequency of higher excited modes also decreases as $V_0$ approaches zero~\cite{bristy_25}. In the inset of Fig.~\ref{mode_with_V}, we show the dependence of the dipole mode frequency on $V_0$ for different values of spatial width $\sigma$. For an attractive potential well, the dipole mode frequency is higher for larger $\sigma$, as a wider potential promotes the center-of-mass motion of droplets, leading to a higher restoring force. In contrast, the weaker confinement with smaller $\sigma$ results in a reduced frequency of dipole mode.

As discussed in Fig.~\ref{density_w_natoms}, the ground state of a self-trapped state is strongly influenced by the potential defect and the number of atoms $N$. The low-lying excitation spectrum as a function of $N$ for three values of potential strengths, $V_0=-0.5$, $0$, and $0.3$ are shown in Fig.~\ref{mode_w_natoms}. We primarily highlight the evolution of dipole and breathing modes. With an increase in $N$, the interaction effects become more pronounced and the condensate localizes near the center of the trap [cf. Fig.~\ref{density_w_natoms}(a)]. As a result, the dipole mode softens to zero energy beyond a critical atom number ($N_{\text{cr}} \gtrsim 150$). The critical $N$ decreases as the $g_{12}$ becomes less attractive. This soft mode arises from interaction-induced modifications of the ground state of the system, where the formation of a strongly localized central peak allows low-energy internal density oscillations around the condensate core. The flatter region of the potential near the minima decreases the effective curvature of the potential, and thus results in weaker restoring force relative to the droplet's inertia. The frequency of the breathing mode also decreases with increasing interaction caused by larger $N$.  

Next, we show the frequencies of collective modes in the absence of a defect in Fig.~\ref{mode_w_natoms}(b). Such a system without any potential defect has previously been studied recently in Refs.~\cite{du_23,fei_24,charalampidis_25}. In the absence of an external confinement, the system exhibits translational invariance, resulting in a zero-energy dipole mode (center-of-mass)  corresponding to a rigid translation of the condensate without altering its internal structure. This is evident from the green curve, which coincides with the zero-energy branch and is consistent with Fig.~\ref{mode_with_V}. As the breathing mode corresponds to a compressional oscillation of the condensate size, the interaction effect (due to larger $N$) modifies the equilibrium ground state density and compressibility of the condensate, leading to a gradual lowering in the frequency of breathing mode.
\begin{figure}[ht]
  \includegraphics[width=0.5\textwidth]{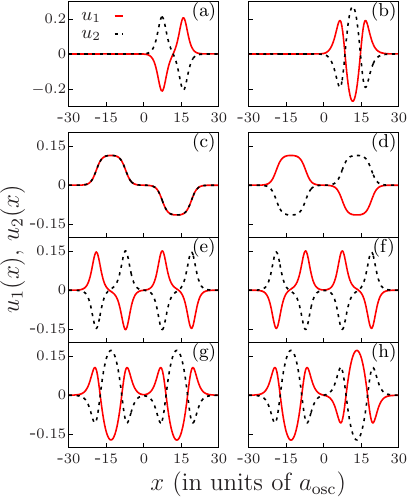}
  \caption{Quasiparticle amplitudes $u_1(x)$ and $u_2(x)$ of the collective excitations of two-component droplets. (a) and (b) correspond to the polarized state with $N=50$, showing the 
           dipole mode and breathing mode, respectively. (c)–(h) correspond to the unpolarized state with $N=150$. Plots (c) and (d) show the dipole modes corresponding to in-phase and out-of-phase oscillations of the two components, respectively. Plots (e) and (f) illustrate the symmetric double-well dipole mode (in-phase) and the asymmetric double-well dipole mode (out-of-phase). (g) and (h) represent the symmetric and asymmetric breathing modes of the system. The spatial coordinate $x$ is expressed in units of the oscillator length $a_{\mathrm{osc}}$.}
  \label{quasi_amplitude}
\end{figure}
\begin{figure}[h]
  \includegraphics[width=0.5\textwidth]{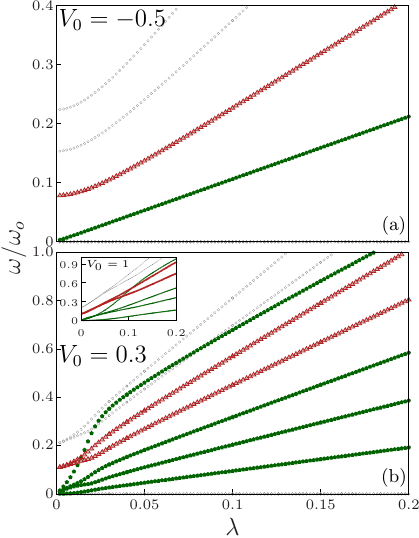}
  \caption{Low-lying collective mode frequencies as a function of the trapping strength $\lambda$ for $N = 200$. Upper panel (a) corresponds to $V_0 = -0.5$, and lower panel (b) corresponds to 
           $V_0 = 0.3$. The inset in panel (b) displays the excitation spectrum for $V_0 = 1$. The green filled stars and red triangles denote the dipole and breathing modes, respectively, while the light-gray circles represent the higher excitation frequencies}
  \label{mode_with_lambda}
\end{figure}

In Fig.~\ref{mode_w_natoms}(c), for the repulsive double-well barrier $V_{0}=0.3$, the equilibrium structure and excitation spectrum are governed by the competition between the repulsive mean-field interaction and the attractive LHY correction. For a small number of atoms ($N \lesssim 10^{2}$), the self-binding energy induced by the LHY correction exceeds the barrier energy scale, resulting in a single localized droplet occupying one side of the double well. In this regime, the excitation spectrum exhibits two zero-energy modes associated with the independent $U(1)$ phase invariance of binary condensate. The dipole mode corresponding to center-of-mass oscillations possesses a small finite energy, which overlaps with zero-energy modes. Unlike harmonic confinement, which pulls the center of mass back to equilibrium, the repulsive barrier breaks translational invariance and allows the self-bound droplet to occupy one of the minima; thus, causes the center of mass oscillations to be very soft. The mode functions corresponding to the dipole and breathing modes in the localized-droplet configuration are shown in Fig.~\ref{quasi_amplitude}(a) and Fig.~\ref{quasi_amplitude}(b). This makes the dipole mode nearly zero-energy mode due to an interplay of barrier repulsion and attractive $g_{12}$. We have explicitly verified that with a less attractive $g_{12}$, the dipole mode deviates from being close to zero-energy. 
\begin{figure*}[ht]
  \includegraphics[width=\linewidth]{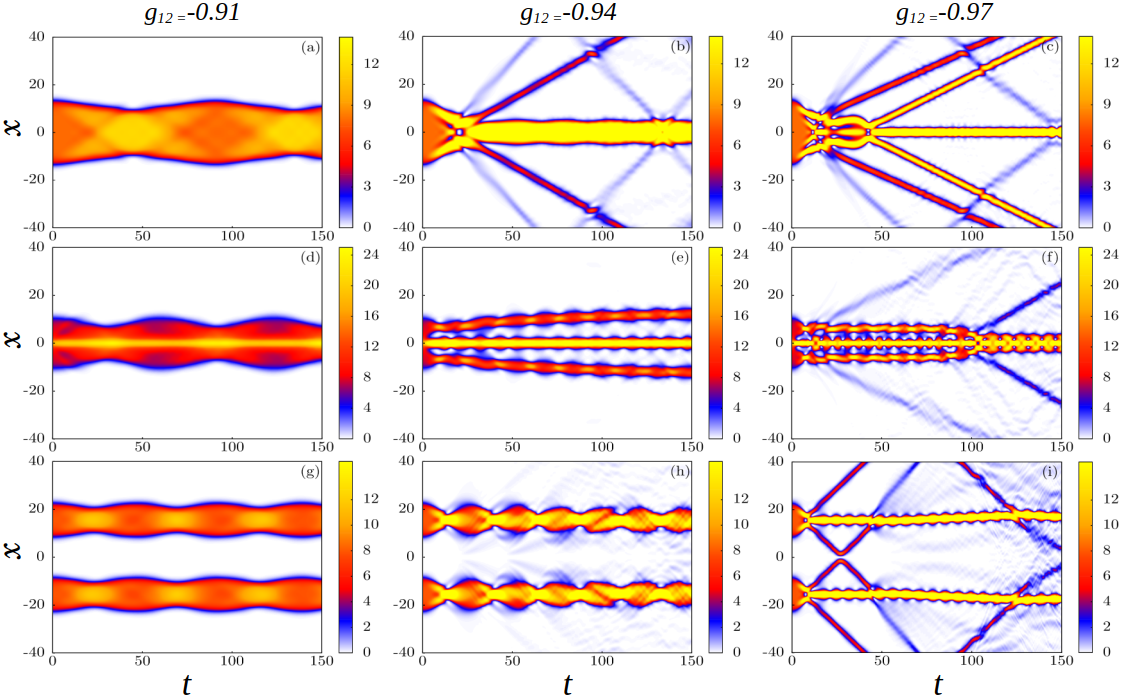}
  \caption{Time evolution of the atomic density of quantum droplet under a sudden quench of intercomponent interaction for different potential defect strengths and intercomponent interactions. The 
           panel correspond to barrier strengths $V_0 = 0$ (top), $V_0 = -0.5$ (middle), and $V_0 = 0.3$ (bottom). The columns represent quenches of $g_{12}$ from the initial value $g_{12}=-0.9$ to $-0.91$ (left), $-0.94$ (middle), and $-0.97$ (right). The figure demonstrates the interplay between barrier strength and intercomponent attraction to control the droplet dynamics, resulting in localization, filament generations, and fragmentation.}
\label{realtime_w_g12}
\end{figure*}

Above a critical $N$, the mean-field interaction energy increases faster than the LHY contribution, effectively reducing the binding relative to the external potential. Consequently, the repulsive barrier energy becomes energetically favourable and the density of ground-state segregates into two fragments, localized in the individual wells. A discontinuity appears in the spectra for the transition from the polarized ground state to the unpolarized droplet state. Here, the number of zero-energy modes increases, as each well confines an individual quantum droplet. In particular, the independent $U(1)$ phase symmetries lead to two zero-energy modes. Moreover, four low-lying excited modes become nearly zero in energy due to the weak coupling between two localized droplets. These modes correspond to (i) the in-phase center-of-mass motion of the two droplets, (ii) the out-of-phase inter-well dipole oscillation, and (iii-iv) symmetric and antisymmetric combinations of localized dipole oscillations within each well, as shown in Fig.~\ref{quasi_amplitude}(c-f). Consequently, the low-energy spectrum consists of six modes clustered around zero energy. In addition, two nearly degenerate breathing modes emerge, representing symmetric and antisymmetric compressional oscillations of the unpolarized configuration, as illustrated in Fig.~\ref{quasi_amplitude}(g–h). 

We further study the role of harmonic confinement in the presence of defect potential. Fig.~\ref{mode_with_lambda} shows the change in mode frequencies for $N=200$ as the harmonic trapping strength $\lambda$ varied. In Fig.~\ref{mode_with_lambda}(a), we consider the case of a central dimple potential with strength $V_0 = -0.5$. For $\lambda = 0$, the system exhibits a zero-energy dipole (additional Goldstone) mode, reflecting the translational invariance of the condensate in the symmetric trap configuration. As $\lambda$ increases, this dipole mode progressively hardens, as shown by a green curve, indicating the restoration of a finite restoring force due to the increase in confinement. Physically, the enhancement of $\lambda$ strengthens the harmonic background, which suppresses the shallow accumulation of central density and gradually flattens the density profile. The breathing mode (shown by the red triangles) also increases monotonically with increasing trapping strength. It is evident that all low-lying excitation branches shift upward with increasing $\lambda$, signaling a transition from a weakly confined, soft collective regime to a more strongly confined, interaction-dominated regime. The increasing separation between the dipole and breathing mode branches further indicates that the system gradually loses its quasi-soft central structure and evolves toward a more rigid, harmonically trapped condensate configuration.

In Fig.~\ref{mode_with_lambda}(b), we consider the repulsive defect case $V_0 = 0.3$, where the double-well structure is more pronounced and the condensate density is distributed in two minima for $N=200$. In this regime, the excitation spectrum becomes richer due to the emergence of multiple nearly zero energy dipole mode oscillations associated with the two wells. For $\lambda = 0$, the spectrum already shows distinct low-lying branches corresponding to different symmetry configurations (mode amplitude is shown in Fig.~\ref{quasi_amplitude}(c-h)). As the trapping strength $\lambda$ increases, four modes progressively harden. The hardening of all four branches of the mode with increasing $\lambda$ reflects the enhanced restoring force provided by a stronger harmonic confinement. As the trap becomes tighter, the tunneling between the wells is reduced, and the effective stiffness of the system increases, thereby increasing the energy cost of center-of-mass and relative oscillations. In addition, breathing modes (denoted red triangles) also shift in frequency with $\lambda$, indicating enhanced compressibility effects under stronger confinement. The increase in the frequency of the breathing mode signifies that the density modulations within each well become energetically more expensive as the system transitions toward a more tightly bound configuration. The inset of panel (b), corresponding to $V_0 = 1$, the stronger repulsive barrier enhances the well separation and requires higher $\lambda$ for the hardening of quasiparticle modes. 

\subsection{Dynamical properties}
We finally examine the nonequilibrium dynamics of the quantum droplet following a sudden quench of the interaction parameter $g_{12}$ for three different barrier strengths. Next, we discuss the dynamical response of the system under both sudden and adiabatic variations of the barrier strength at fixed $g_{12}$.

Figure~\ref{realtime_w_g12} shows the real-time evolution of the droplet density with $N=200$  following a quench of the intercomponent interaction towards stronger attraction, for different barrier strengths. The initial state ($t=0$) is prepared at $g_{12}=-0.9$, and $g_{12}$ suddenly is quenched to more attractive values. The columns of the figure correspond to increasing quench attraction ($-0.91$, $-0.94$, and $-0.97$), while the rows represent different external potentials: no potential ($V_0=0$), an attractive potential ($V_0=-0.5$) and a repulsive potential ($V_0=0.3$). For a homogeneous case with $V_{0}=0$ (top row), the droplet mainly exhibits longitudinal breathing and shape oscillations for weak quench, as shown in Fig.~\ref{realtime_w_g12}(a). As the quench attraction increases, the quench injects excess binding energy into the system, generating strong density waves [see Fig.~\ref{realtime_w_g12}(b)]. For the strongest quench, as depicted in Fig.~\ref{realtime_w_g12}(c), the droplet undergoes pronounced dynamical deformation, indicating the onset of a modulational instability and partial self-focusing driven by the enhanced mean-field attraction competing with the stabilizing LHY repulsion. 

In the presence of an attractive well ($V_0=-0.5$, middle row) as shown in Fig.~\ref{realtime_w_g12}(d,e,f), the confinement provided by the potential localizes the droplet and modifies the collective response. For a smaller quench, the droplet remains trapped and performs a breathing oscillation. At stronger attraction, the density is redistributed into multiple localized filaments aligned with the potential minimum, indicating a transition toward a quasi-bound multi-peak density configuration. For the strongest quench, interference patterns and propagating density fronts appear, signaling energy redistribution between internal collective modes and center-of-mass motion.

For the repulsive barrier ($V_0=0.3$, bottom row) in Fig.~\ref{realtime_w_g12}(g,h,i), the droplet splits into two fragments even for weak quenches, reflecting the competition between self-binding and barrier-induced localization. Increasing the attraction enhances the effective binding of each localized fragment, leading to long-lived oscillatory droplets on either side of the barrier. For the strongest quench, the fragments exhibit strong internal modulation and emit dispersive waves, revealing a regime where the attractive mean-field contribution dominates locally while the LHY correction prevents full collapse, resulting in dynamically stabilized oscillatory structures.
\begin{figure}[htbp]
  \centering
  \includegraphics[width=0.5\textwidth]{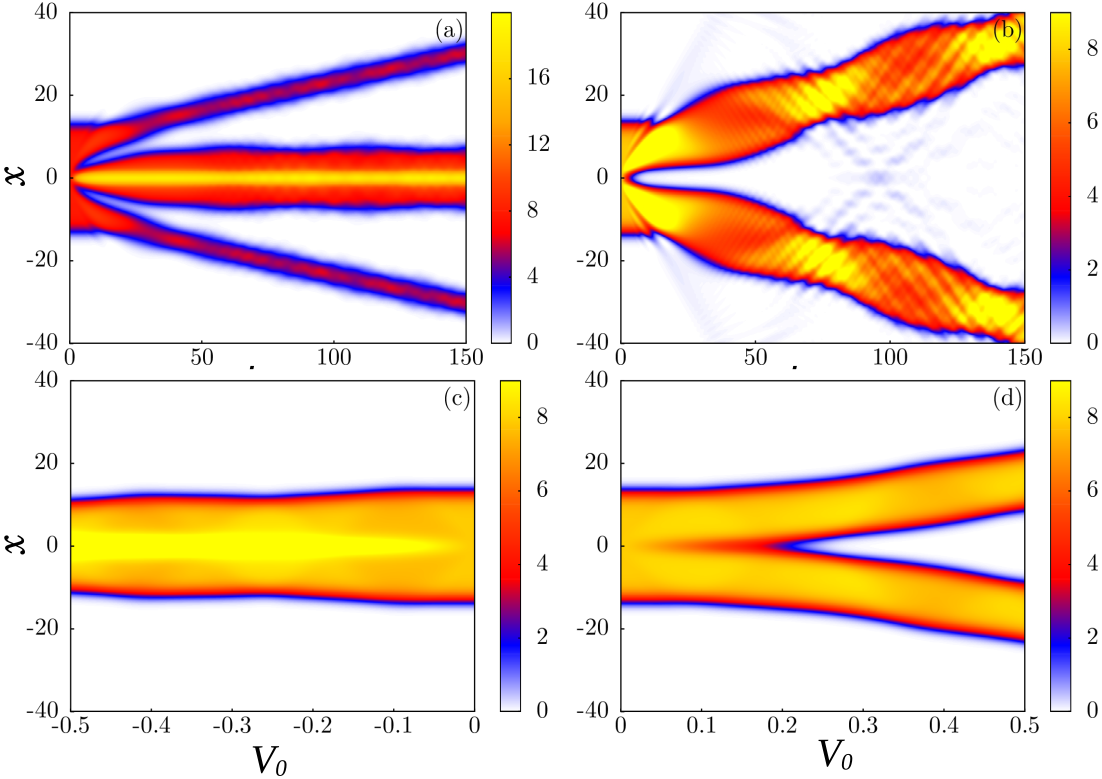}
  \caption{Time evolution of the density of quantum droplet following a quench of the external potential $V_0$ from zero to different values. The initial ground state is prepared with $g = 1$, $N =    
           200$, and $g_{12} = -0.9$. Quenching to weaker attractive or repulsive barrier strengths induces dynamical spreading and fragmentation of droplet (a,c), while stronger quenches lead to droplet splitting (b,d). The upper panel (a,b) corresponds to sudden quenches from $V_0 = 0$ to $-0.5$ and $0.5$, whereas the lower panel (c,d) shows the corresponding adiabatic quenches.}
  \label{realtime_w_u}
\end{figure}

Fig.~\ref{realtime_w_u} illustrates the nonequilibrium response of the quantum droplet to a change in the potential defect $V_0$, starting from the equilibrium state of $V_0=0$ (without defect). 
As the droplet is a self-bound state, modifying the external potential disrupts the balance of $g_{12}$ and LHY and redistributes the internal pressure. For sudden quenches (a,b), the barrier quench injects energy into the system and excites collective modes. When the barrier is quenched to a weak attractive value [Fig.~\ref{realtime_w_u}(a)], the droplet experiences an effective longitudinal compression, leading to fragmentation into three droplets by emission of dispersive waves. In contrast, a sudden quench to a repulsive barrier [Fig.~\ref{realtime_w_u}(b)] creates an outward pressure at the center, producing a density depletion and initiating the separation of the droplet into two propagating fragments. The resulting dynamics are characterized by strong interference patterns and density oscillations, indicating significant excitation of internal breathing and dipole modes.

For adiabatic ramps (c,d), the system remains closer to instantaneous equilibrium and therefore exhibits steady evolution.  A gradual ramp toward an attractive barrier (c) primarily modifies the droplet width without generating strong excitations, showing that the droplet adapts by redistributing its density to minimize the combined interaction and potential energy. However, an adiabatic increase toward a repulsive barrier (d) leads to a controlled splitting of the droplet into two self-bound fragments, with minimal emission of radiations. This behavior reflects the ability of the droplet to reorganize its internal pressure balance when the external perturbation is slow compared to the intrinsic timescale of collective excitations. 

\section{Conclusions}\label{conclusion}
We have investigated the stationary properties, collective excitations, and nonequilibrium quantum dynamics of two-component quantum droplets under a potential defect. Our study is based on the
numerical simulation of the extended Gross-Pitaevskii equation that incorporates the beyond mean-field exact LHY correction. The exact treatment allows us to examine the collective excitations in the wider (away from collapse threshold) attractive interaction and atomic density regime. The number of atoms, the characteristic of the external defect, and the introduction of a harmonic trap result in intriguing collective behaviour of many-body droplet mixtures. The number of atoms drives a transition from the polarized to an unpolarized state in the presence of a repulsive barrier, which is accompanied by a discontinuity in the collective excitation spectrum of low-lying modes. As the attractive interaction overcomes the tendency of the barrier to separate the components, the unpolarized state becomes energetically favourable at a smaller number of atoms for larger intercomponent interactions. In the presence of an attractive defect, the flattening of the potential near the minima results in softening of the dipole mode with the number of atoms, which gains in energy as harmonic confinement is introduced. The weak quench of attractive interaction leads to breathing oscillations, whereas strong quench causes the generation of filaments and fragmentation into multi-droplet states at long-time evolution. Breakup of droplet mixtures also occurs for quenching the attractive or repulsive defect potential. Our study presents the polarization transition and collective spectra that could be useful for the recent ultracold experiments on quantum droplet and Josephson dynamics.    

\begin{acknowledgments}
We thank Sandeep Gautam, Rajat, and Ritu for the valuable discussions. H.K. acknowledges financial support from the University Grant Commission (UGC), New Delhi. K.S. acknowledges support from the Science and Engineering Research Board, Department of Science and Technology, Government of India through Project No. SRG/2023/001569.
\end{acknowledgments}

\bibliographystyle{apsrev4-1} 
\bibliography{reference}
\end{document}